\documentclass[12pt,a4paper]{article}
\usepackage{amsmath}
\usepackage{graphicx}

\begin{document}
\title{Discrete Self-Similarity in Type-II Strong Explosions}
\author{Yonatan Oren \\ \small{Racah Institute of Physics, the Hebrew University, 91904 Jerusalem, Israel} \\
       Re'em Sari \\ \small{ Racah Institute of Physics, the Hebrew University, 91904 Jerusalem, Israel} 
                  \\ \small{and California Institute of Technology, MC 130-33, Pasadena, CA 91125 }}
\date{}                  

\maketitle

\begin{abstract}
We present new solutions to the strong explosion problem in a non-power law density profile. The unperturbed self-similar solutions discovered by Waxman \& Shvarts describe strong Newtonian shocks propagating into a cold gas with a density profile falling off as $r^{-\omega}$, where $\omega>3$ (Type-II solutions). The perturbations we consider are spherically symmetric and log-periodic with respect to the radius.  While the unperturbed solutions are continuously self-similar, the log-periodicity of the density perturbations leads to a discrete self-similarity of the perturbations, i.e. the solution repeats itself up to a scaling at discrete time intervals. We discuss these solutions and verify them against numerical integrations of the time dependent hydrodynamic equations. Finally we show that this method can be generalized to treat any small, spherically symmetric density perturbation by employing Fourier decomposition.
\end{abstract}

\section{Introduction}
Expanding shock waves are naturally produced by diverse astrophysical phenomena, such as supernovae, GRBs, stellar winds and more. So far, analytical self-similar solutions have been found for several simple cases, of  which we take special interest in the case of strong spherical shocks propagating into a density profile that decays as a power of the radius, $\rho=K r^{-\omega}$.
The first solutions of this kind to be found, now commonly known as the Sedov-Taylor (ST) solution, were given by Sedov, Taylor and Von-Neumann \cite{Sedov} \cite{Taylor} \cite{VonNeumann} for the case $\omega<3$ to describe decelerating shocks. In this paper however we shall consider a second class of solutions which was discovered by Waxman and Shvarts (WS) for the $\omega>3$ case \cite{WS}. In these solutions the shock front accelerates because of the rapid decay of the density ahead of the shock, causing a part of the flow to be causally disconnected from the inner region containing the source of the explosion. Mathematically, the boundary  of this region appears as a singular point of the hydrodynamic equations somewhere between the explosion and the shock, called the sonic point.  
 
The solutions discussed above, while useful, fall short when describing shocks propagating into density profiles that deviate from a simple power law decay. This might occur in a variety of astrophysical scenarios, e.g. a supernova shock propagating into a modulated stellar wind. For this reason it is desirable to generalize as much as possible the external density profile for which we can obtain analytic solutions, and this is what we attempt here. 
It should be clarified that while we deal with perturbations we do not perform an  analysis of stability, but only find solutions corresponding to perturbed external conditions. The stability of First type solutions has been studied by Ryu \& Vishniac \cite{ryu} and Kushinr et al. \cite{Kushnir}, and that of second type solution by Sari et al. \cite{SWS}, and much of the formalism used for the perturbative analysis in this paper is similar these works. 

The plan of the paper is as follows: In section \ref{sec:unpert} we review the unperturbed solutions and the boundary conditions at the shock front and at the sonic point. In section \ref{sec:pert} we develop the perturbation equations and boundary conditions. We then discuss the solutions to these equations and compare them to numerical results obtained from a full hydrodynamic treatment of the problem. In section \ref{sec:fourier} we discuss a method of generalizing our results to accommodate arbitrary small density perturbations, and finally conclude in section \ref{sec:discussion}.

\section{The Unperturbed Solution} \label{sec:unpert}
We proceed to give a quick review of the unperturbed solutions under consideration \cite{WS}. The physical scenario is the discharge of a large amount of energy from a point source at the center of a spherically symmetric distribution of cold gas. It may be noted that spherical symmetry was chosen for its relevance to most astrophysical scenarios, but planar and cylindrical geometries may readily be treatd as well. The gas density follows a power law behavior such that $\rho (r)=K r^{-\omega}$. 
\subsection{The hydrodynamic equations}  
We begin with the Euler equations for an ideal fluid with adiabatic index $\gamma$ in spherical symmetry:
\begin{align}
(\partial_t +u \partial_r) \rho + \rho r^{-2} \partial_r (r^2 u) &= 0 \nonumber\\
\rho (\partial_t +u \partial_r) u + \partial_r (\gamma^{-1} \rho c^2) &= 0 \nonumber\\
(\partial_t +u \partial_r) (\gamma^{-1} c^2 \rho^{1-\gamma}) &= 0.
\label{eq:hydro}
\end{align}
These equations feature the density $\rho$, velocity $u$ and speed of sound $c$ as the dependant variables, while the pressure has been eliminated through the equation of state $p=\gamma^{-1} \rho c^2$. We will use $p$ or $c$ interchangeably as the third dependant variable by merit of convenience. The only relevant scale in the problem at late times is the shock radius $R(t)$, and so the self similar solutions are given in terms of the variable $\xi=r/R(t)$. We define the self-similar functions $U$,$C$,$G$ and $P$ such that the solutions take this form:
\begin{align}
u(r,t) &= \dot{R}\xi U(\xi) \nonumber\\
c(r,t) &= \dot{R}\xi C(\xi) \nonumber\\
\rho(r,t) &= B R^{-\omega} G(\xi) \nonumber\\
p(r,t) &= B R^{-\omega} \dot{R}^2 P(\xi) .
\label{eq:SSdefs}
\end{align}
These definitions are supplemented by a scaling law for the shock radius, $\dot{R} \propto R^\delta$. This determines the time dependence of the radius to be 
\begin{align}
R(t) = \left\{ \begin{array}{rl}
 A(t-t_0)^\alpha, \delta<1 \\
 A e^{t/\tau}, \delta=1 \\
 A(t_0-t)^\alpha, \delta<1, 
       \end{array} \right.
\label{eq:Roft}
\end{align}
where $\alpha=1/(1-\delta)$. The fixing of $\delta$ will be discussed shortly.
Substituting Eq.\ref{eq:SSdefs} into Eq.\ref{eq:hydro} we obtain two equations for the self similar functions $U$ and $C$:
\begin{align}
\frac{dU}{d log\xi} &= \frac{\Delta_1(U,C)}{\Delta(U,C)} \label{eq:SSU}\\
\frac{dC}{d log\xi} &= \frac{\Delta_2(U,C)}{\Delta(U,C)},\label{eq:SSC}
\end{align}
where the functions $\Delta$, $\Delta_1$ and $\Delta_2$ are
\begin{align}
\Delta &=C^2-(1-U)^2 \nonumber\\
\Delta_1 &= U(1-U) \left( 1-U-\frac{\alpha-1}{\alpha} \right)
           -C^2 \left( 3U-\frac{\omega-2[(\alpha-1)/\alpha]}{\gamma} \right) \nonumber\\
\Delta_2 &= C \left\{ (1-U) \left( 1-U-\frac{\alpha-1}{\alpha} \right) 
           - \frac{\gamma-1}{2}U \left( 2(1-U)+\frac{\alpha-1}{\alpha} \right) \right. \nonumber\\      	&\left. - C^2 +\frac{C^2}{1-U}\frac{(\gamma-1)\omega+2[(\alpha-1)/\alpha]}{2\gamma} \right\},
\end{align}
and an implicit condition for $G$:
\begin{equation}
C^{-2}(1-U)^\lambda G^{\gamma-1+\lambda}\xi^{3\lambda-2}=Constant
\end{equation}
where 
\begin{equation}
\lambda=\frac{2\delta+(\gamma-1)\omega}{3-\omega}.
\end{equation}
Evidently for the solution to pass smoothly through a singular point there must exist some $\xi_s$ where \begin{equation}
\Delta=\Delta_1=\Delta_2=0.
\label{eq:sonicpoint}
\end{equation}

\subsection{Boundary conditions}
The boundary conditions at the shock front are determined by the Rankine-Hugoniot conditions \cite{Landau} applied to a strong shock. In terms of the self-similar functions these turn out to be
\begin{align}
U(\xi=1) &= \frac{2}{\gamma+1} \nonumber\\
C(\xi=1) &= \frac{\sqrt{2\gamma(\gamma-1)}}{\gamma+1} \nonumber\\
G(\xi=1) &= \frac{\gamma+1}{\gamma-1}.
\label{eq:BCshock}
\end{align}
The value of $\delta$ has yet to be determined, and to find it we need an additional condition.
This condition is supplied by the requirement that the solution pass smoothly through the sonic point, namely that Eq.\ref{eq:sonicpoint} has a solution. Solving this system yields the values of $\xi_s$ and $\delta$, completing the solution in the unperturbed case. In general these values can only be found numerically.  
Type-II solutions with a sonic point have been found and are believed to exist for $\omega>\omega_g(\gamma)>3$. There is a small range of $3<\omega<\omega_g(\gamma)$ between the first and second type solutions where a third kind of self similar solution exists \cite{Gruzinov}, which are out of the scope of this paper.

\section{Discretely self similar perturbations}\label{sec:pert}

\subsection{The perturbation equations}
We now come to the case of a perturbed density profile. For the perturbation equations to be tractable we aim at a self similar solution by carefully choosing a perturbation whose characteristic wavelength scales like radius. Namely, we take the perturbed density profile to be 
\begin{equation}
\rho(r)+\delta \rho (r)=K r^{-\omega} \left[ 1+\sigma \left(\frac{r}{r_0}\right)^{i\beta} \right]
\end{equation}
where $r_0$ has dimensions of length and bears only on the phase of the perturbation. $\rho(r)$ is the unperturbed density, $\beta$ is the frequency of the perturbation and $\sigma$ is a small, real, and dimensionless amplitude. Here and elsewhere we take the real part of any complex quantity to be the physically significant element.

The perturbed solution is defined as
\begin{align}
u(r,t)+\delta u(r,t) &= \dot{R}\xi [U(\xi)+f(t) \delta U(\xi)] \nonumber\\
\rho(r,t)+\delta \rho(r,t) &= K R^{-\omega} [G(\xi)+f(t) \delta G(\xi)] \nonumber\\
p(r,t)+\delta p(r,t) &= K R^{-\omega} \dot{R}^2 [P(\xi)+f(t) \delta P(\xi)],
\label{eq:SSpertdef}
\end{align}
and
\begin{equation}
R(t)+\delta R(t)=R(t) [1+f(t)].
\label{eq:Rpertdef}
\end{equation}
To enable separation of variables, the function $f(t)$ is taken to obey 
\begin{equation}
f(t)=\frac{\sigma}{d} \left(\frac{R}{r_0}\right)^q \Rightarrow \frac{\dot{f}R}{f\dot{R}}=q,
\end{equation}
where $q$ describes the frequency and $d$ the amplitude and phase of the perturbations behind the shock. Finding the values of these parameters is discussed in section \ref{sec:pertbc}. 

Plugging Eq.\ref{eq:SSpertdef} into Eq.\ref{eq:hydro} and taking the first order in $f(t)$ yields the self-similar linear equations for the perturbations. This set of equations can be written as 
\begin{equation}
M Y'=L Y
\end{equation}
where 
\begin{align}
Y(\xi)=\begin{bmatrix} \delta G \\ \delta U \\ \delta P \end{bmatrix},
M=\begin{bmatrix} 
  	(U-1)\xi				& G\xi 		 	& 0 				\\
  	0						& G(U-1)\xi^2   & 1 				\\
    -\gamma \xi \frac{U-1}{G}	& 0				& \xi \frac{U-1}{P} 	
  \end{bmatrix}, \nonumber \\
L=-\begin{bmatrix} 
  	q-\omega+3U+\xi U'						& 3G+\xi G'	 					& 0				\\ \\
  	-\frac{P'}{G}							& \xi G(q+\delta -1+2U+\xi U')   & 0				\\ \\
    \gamma \xi \frac{(U-1)G'}{G^2}-\gamma \frac{q}{G}	& \xi(\frac{P'}{P}-\gamma \frac{G'}{G})	& \frac{q}{P}-\xi \frac{(U-1)P'}{ P^2} 	
  \end{bmatrix}
\end{align}

\subsection{Boundary conditions for the perturbations}\label{sec:pertbc}
At the shock front, the perturbed solution must also obey the Hugoniot conditions. From this requirement we find that $q=i\beta$ and that the boundary conditions for the perturbations are
\begin{align}
\delta G(\xi=1) &= \frac{\gamma+1}{\gamma-1}(d-\omega)-G'(1) \nonumber \\
\delta U(\xi=1) &= \frac{2}{\gamma+1}q-U'(1) \nonumber \\
\delta P(\xi=1) &= \frac{2}{\gamma+1}[2(q+1)-\omega+d]-P'(1).
\label{eq:BCshockpert}
\end{align}
In analogy to the unperturbed solution, where the parameter $\delta$ was fixed by requiring the solution to pass smoothly through the sonic point, we here fix $d$ by the same requirement for the perturbation functions. For a general value of $d$, if we start with Eq.\ref{eq:BCshockpert} and solve back towards decreasing radii the solution will diverge at the sonic point. There is, however, one value of $d$ where this will not happen, and that is the physical value that we seek.

We can now see that since the real part of $f(t)$ is periodic, the solution is discretely self-similar, i.e. it repeats itself up to a scaling factor in intervals of $\frac{\Delta R}{R}=e^{\frac{2\pi}{\beta}}-1$. While the unperturbed solution and the perturbations in their complex form are both self-similar, the physical solution which is the real part of their sum is not.   

\subsection{The Discretely self-similar solution}
While self similarity simplifies the problem by reducing the partial differential equations for the perturbations into ordinary differential equations, still these are generally not analytically solvable. Therefore, for each specific set of values for $\gamma$, $\omega$ and $\beta$ we must find numerically the functions $\delta G$,$\delta U$ and $\delta P$ and the parameter $d$. This can be straightforwardly done, once $d$ is found, by integrating back from the shock towards the sonic point.

\begin{figure}
\includegraphics[scale=0.8]{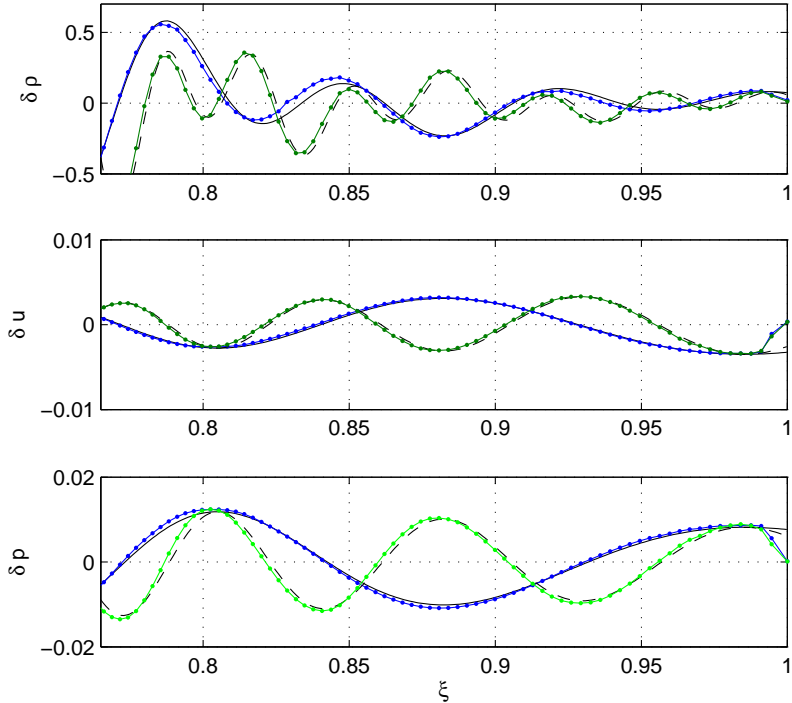}
\caption{The self-similar functions $\delta G$, $\delta U$ and $\delta P$ as calculated in a numerical simulation with $\gamma=5/3$, $\omega=17/4$ and $\beta=20$ and $40$ for the blue and green lines respectively. The solid and dashed black lines are the real parts of the corresponding solutions of the perturbation equations for $\beta=20$ and $\beta=40$ respectively.}
\label{fig:GUPalpha}
\end{figure}

\begin{figure}
\includegraphics[scale=0.6]{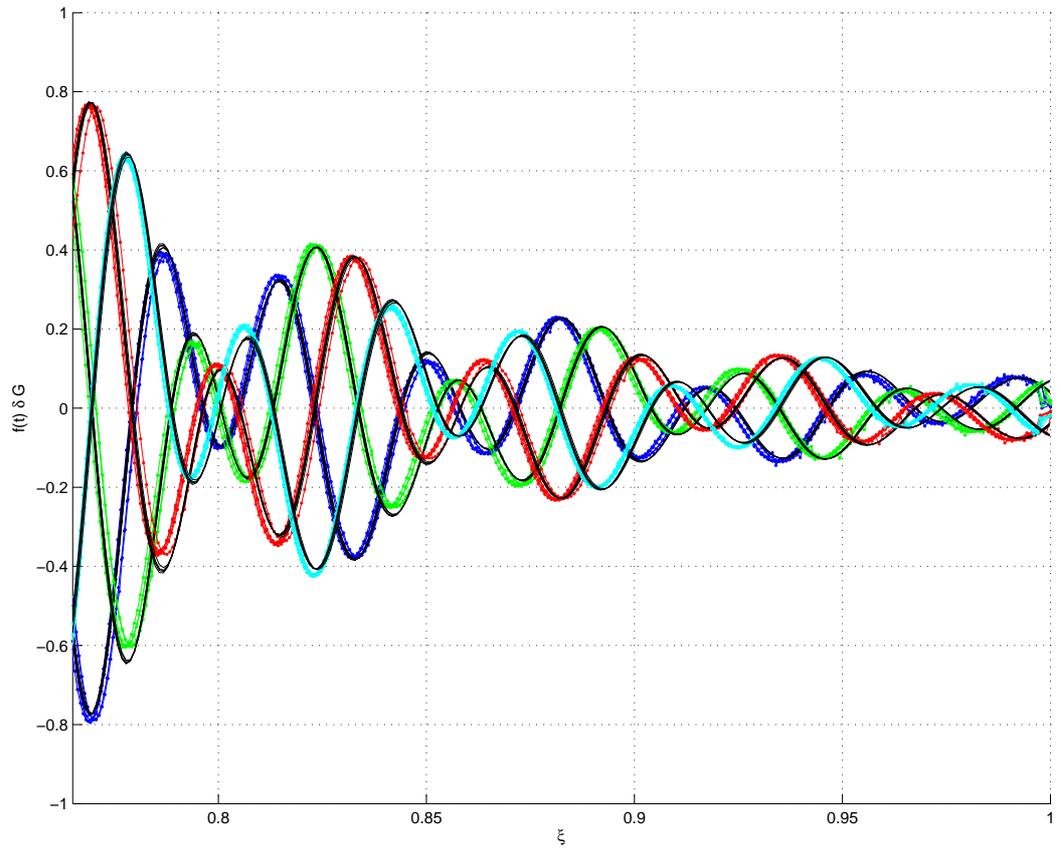}
\caption{Discrete Self-similarity: $\delta G$ is plotted in four different phases of its periodical repetition. The blue, green, red and cyan lines correspond to a phase shift of $0$,$1/4$,$1/2$ and $3/4$ period relative to figure \ref{fig:GUPalpha}. In each color five different periods are drawn, but because of the discrete self-similarity they overlap and are almost indistinguishable.}
\label{fig:DSS}
\end{figure}

In figure \ref{fig:GUPalpha} we present some solutions to the perturbation equations discussed above, against a numerical solution of the partial differential hydrodynamic equations given by Eq.\eqref{eq:hydro}, corresponding to the same external density profile and adiabatic index. The code we employ uses a second order Godunov scheme to numerically evolve the hydrodynamic equations. The two solutions are identical up to small numerical errors, verifying the validity of our method. In figure \ref{fig:DSS} the solution for $\delta G$ is plotted at 20 different times, separated by a quarter of the period of the density perturbation. Thus five periods of the perturbations are represented, and four different phases within each. Clearly the shape of the physical solution as a function of $\xi$ changes with time because of the factor $R^{i\beta}$ in the function $f(t)$, and yet it repeats itself at discrete periods, making it discretely, rather than continuously, self-similar.  

It stands out in figure \ref{fig:GUPalpha} that the characteristic wavelength of the density perturbations is notably shorter than that of the velocity and pressure perturbations. This happens because the density supports both traveling sound waves and stationary (in the fluid rest frame) fluctuations for which the pressure and velocity are not perturbed, while the pressure and velocity perturbations must propagate as travelling waves. The dominant component of these perturbations near the shock  are left traveling sound waves, due to the relative weakness of reflected waves from the hydrodynamic profiles behind the shock. From this argument it follows that the characteristic wavelengths are given by $\frac{2\pi}{\beta}(1-\xi U+\sqrt{\gamma \frac{P}{G}})$ for the pressure and velocity, together with $\frac{2\pi}{\beta}(1-\xi U)$ for density perturbations.

\begin{figure}
\includegraphics[scale=0.6]{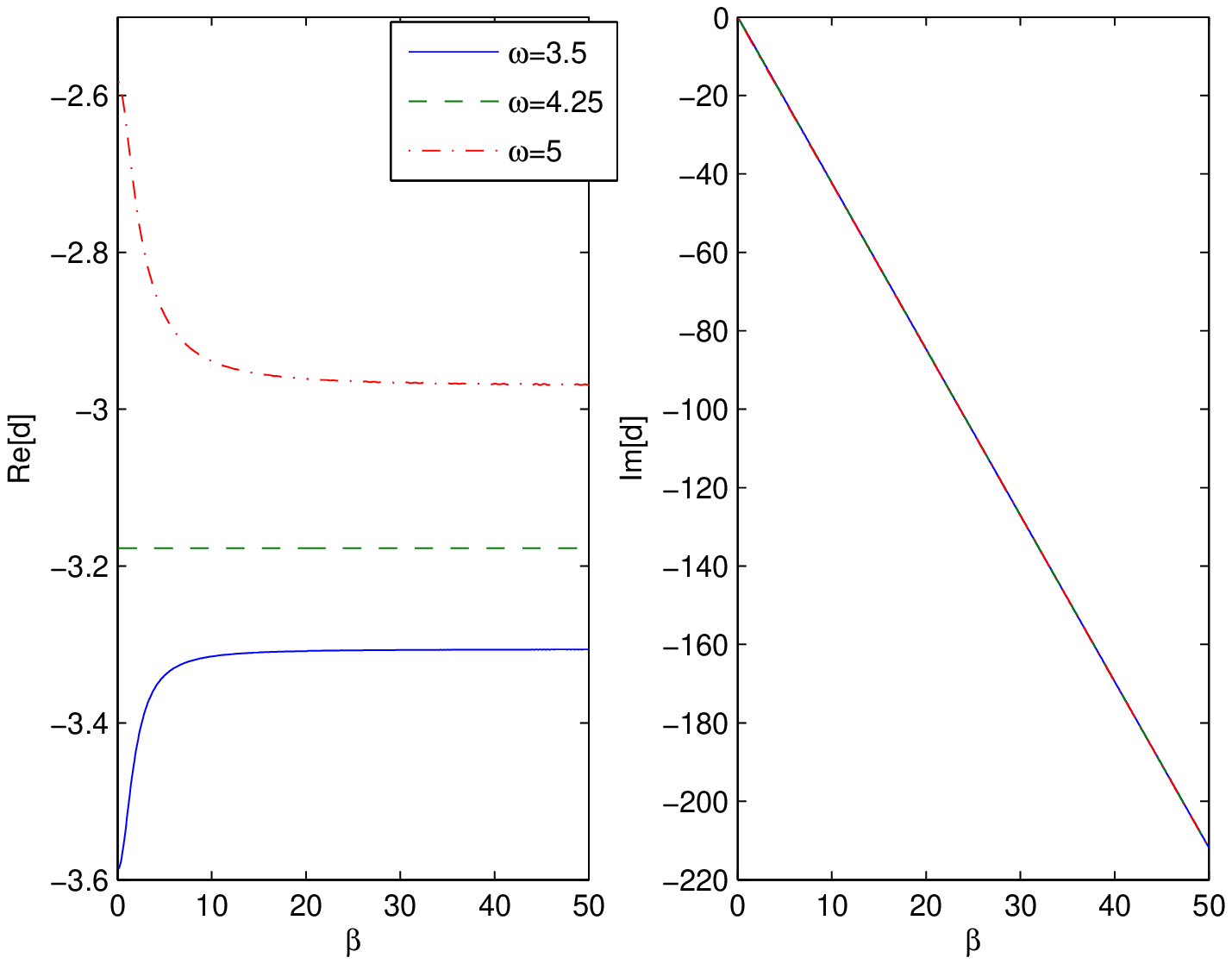}
\caption{The real and imaginary parts of $d(\beta)$ for several values of $\omega$ at $\gamma=\frac{5}{3}$.}
\label{fig:dofag}
\end{figure}

\begin{figure}
\includegraphics[scale=0.6]{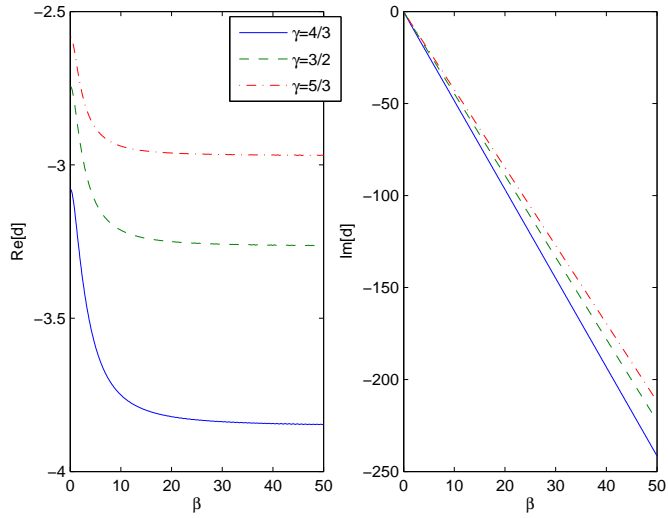}
\caption{The real and imaginary parts of $d(\beta)$ for several values of $\gamma$ at $\omega=5$. The slopes of $Im(d)$ computed numerically from this graph are $(4.830,4.450,4.237)$ for $\gamma=(4/3,3/2,5/3)$, and agree to 3 decimal places with those predicted by equation. \eqref{eq:dhighq} for the respective values of $\gamma$.}
\label{fig:dofaw}
\end{figure}

Finally in figures \ref{fig:dofag} and \ref{fig:dofaw} we look at the parameter $d(\beta)$, relating the fractional perturbation in the shock position to the fractional perturbation in the external density, for several values of $\omega$ and $\gamma$. It can be seen that while the real part of $d$ is roughly constant and of order unity, the imaginary part grows approximately linearly with $\beta$. The implication for the perturbations is that $\delta R$ becomes small (on the order of $\sigma/|d|$) when $\beta$ is large. This is physically sensible since when $\beta \rightarrow \infty$ the perturbations oscillate so quickly that the shock position, which is the integral of the shock velocity, does not respond quickly enough to be significantly affected by the perturbations. The other perturbations, $\delta G$, $\delta U$ and $\delta P$, are themselves of order $d$, as can be seen from Eq.\eqref{eq:BCshockpert}, and so the actual perturbations to the hydrodynamic quantities remain of order $\sigma$. 
We can work out the value of the imaginary part of $d$ when $\beta$ tends to infinity by considering the short wavelength limit. The wavelength of the wave excited by the external density perturbations is then very short compared to the scale of variations in the background hydrodynamic quantities, and these perturbations can then be approximately treated as left traveling waves in a uniform medium. Such waves satisfy $\delta u = - \delta p / (\rho c)$ \cite{Landau}, and using equation. \eqref{eq:BCshock}, \eqref{eq:BCshockpert} we arrive at the relation 
\begin{equation}\label{eq:dhighq}
d \rightarrow -\left( 2+\sqrt{\frac{2\gamma}{\gamma-1}} \right)i\beta,
\end{equation}
which for large $q$ gives an excellent approximation to the numerically calculated value for $Im(d)$, explaining both the magnitude and phase of high frequency perturbations. We defer discussion of the small wavenumber limit to Appendix \ref{app:shifted}.


\section{Arbitrary perturbations} 
\label{sec:fourier}
We have so far discussed the solutions of the perturbation equations for small log-periodic perturbations in the external density. However, the linearity of the perturbations allows us to construct more general solutions by a Fourier decomposition of any periodic spherically symmetric density perturbation. We thus treat our basic solutions as a basis for a more general solution space. It should be noted that while the basic perturbations are self similar (discretely so considering only the real part, but continuously if treated as complex functions), a solution that is the sum of solutions with different $\beta$'s will not be self similar and will have a time dependent profile. This can be plainly understood by analogy to the Schroedinger equation in quantum mechanics. There, each energy eigenfunction is time independent up to a phase, but once different eigenfunctions are combined, their sum becomes time dependent owing to the phases of different eigenfunctions changing at a different rate. 

\begin{figure}
\includegraphics[scale=0.6]{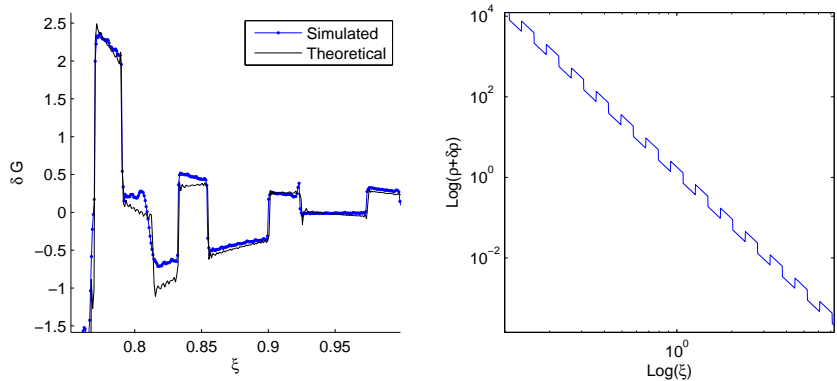}
\caption{Left: Theoretical and numerical results for a square wave density perturbation with $\beta=20$ and $\sigma=0.1$. Right: The power law external density profile perturbed by a square wave, plotted with $\sigma=0.8$ to make the perturbations more visible.}
\label{fig:fourier}
\end{figure}

We confirm the validity of this method by solving the full nonlinear PDEs numerically for a square wave density perturbation. In figure \ref{fig:fourier} we compare this solution to the theoretical solution obtained by summing a large, but finite, amount of the Fourier components that constitute the desired external density profile, where each component is calculated using the methods described above. This cut off series inevitably creates unphysical oscillations (Gibbs phenomenon) in the theoretical solution which may be disregarded. 

This method can in principle be applied even to non-periodic perturbations, such as an isolated pulse or step function (density jump). These would require a continuous Fourier transform to accomplish which is technically difficult when, as is the case here, the perturbations are obtained by numerically solving an ODE. 

\section{Discussion}\label{sec:discussion}
We have laid out a method for solving the strong explosion problem in density profiles that deviate from a pure power law radial dependence. The key lies in choosing radially log-periodic perturbations which do not introduce a new scale into the problem. This leads to self-similar perturbation in the hydrodynamic quantities behind the shock, which can be found by solving a set of ordinary differential equations. The perturbations are fully self-similar when the density perturbation is formally taken to be a complex function,  $\delta \rho / \rho=\sigma \left(\frac{r}{r_0}\right)^{i\beta}$, but  taking the physical real part of the solution makes the perturbations, as well as the full solution (the sum of the unperturbed solution and the perturbation), only discretely self similar because of the periodic nature of the perturbation. We find that the coefficient $d$ connecting the amplitude of the perturbations in the shock position with the amplitude of the density perturbations has a $O(1)$ real part and an $O(\beta)$ imaginary part, so at short wavelengths, $\beta \gg 1$, this perturbation becomes small and is at a relative phase of a quarter wave behind the density perturbation.

\begin{figure}
\includegraphics[scale=0.6]{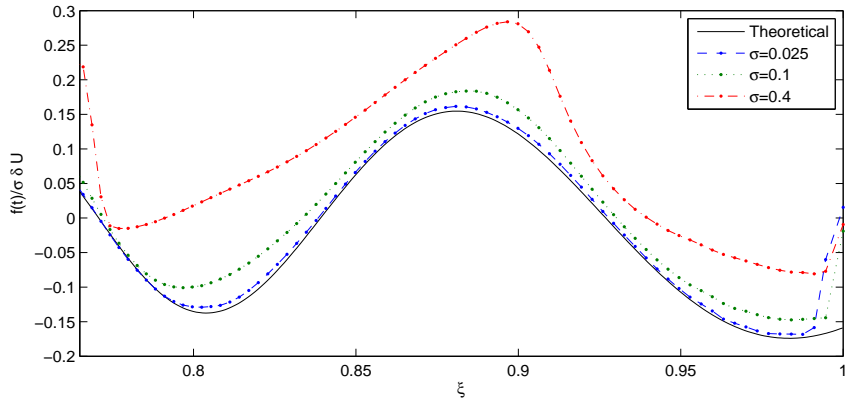}
\caption{Numerical simulation results for the velocity perturbation at $\beta=20$, $\gamma=5/3$, $\omega=17/4$ and different values of $\sigma$, divided by $\sigma$. The difference between the lines comes from nonlinear terms that become pronounced as $\sigma$ increases. Small values of $\sigma$ show a good agreement of the numerical simulation to the linear approximation of our method.}
\label{fig:sigmas}
\end{figure}

The linearized perturbation treatment naturally ensures that the perturbations will not depend on $\sigma$ other than by a linear scaling of the amplitude. This simplifies the solution of the problem but limits the validity of the method to small perturbations. It can be seen in figure \ref{fig:sigmas}, where the solution for a specific perturbation is plotted for several different values of $\sigma$,  that this limitation becomes pronounced when $\sigma \approx 0.1$, although it gives qualitatively correct results even for much higher values. When $\sigma$ is small the difference between the simulated and exact values is quadratic with $\sigma$, which is manifested in figure \ref{fig:sigmas} as a linear scaling due to the $\sigma$ normalization of the different plots.

A natural extension of the argument presented here is to cover the relativistic regime. The basis for such a study would be self-similar solutions for power law density profiles. These were discovered for the ultra relativistic limit by Blandford \& McKee \cite{BM} for first type solutions with $\omega<4$, and by Best \& Sari \cite{best} for second type solutions with $\omega>5-\sqrt{3/4}$. An exploration of relativistic similarity solutions in various geometries (planar, cylindrical and spherical) was given by Sari \cite{sari}. Additionally Pan \& Sari \cite{pan} studied the case where a shock traverses a star's interior and then emerges into empty space. All of these works are valid starting points for perturbative analyses such as the one presented here, and will possibly be pursued in future work.

Acknowledgments: The authors wish to thank Prof. Tsvi Piran for fruitful discussions. This research was partially 
supported by a NASA grant, IRG grant and a Packard Fellowship.

\appendix
\section{Appendix: The small wavenumber limit}
\label{app:shifted}
In considering the family of solutions presented above, an interesting limiting case presents itself in the form of the long wavelength limit, namely when the frequency $\beta$ vanishes. In this limit the flow at any stage will converge to a self similar form before the density perturbation changes significantly. In other words, the solution at any instant should be an unperturbed solution of the type discussed in section \ref{sec:unpert}, but with the magnitude $K$ of density profile slowly changing with time. The value of $K$ is still not enough to uniquely determine the form of the solution, since it is possible that when $K$ changes the parameters $A$ and $t_0$ (see equation. \eqref{eq:Roft}) also change, expressing a change of the energy or starting time of the explosion. It can be shown that if we take $\delta A / A=\sigma/d$ and $\delta t_0=0$ (while from its definition $\delta K/K=\sigma$) we obtain a solution of the form
\begin{align}
\delta G(\xi) &= (d-\omega)G(\xi)-\xi G'(\xi) \nonumber \\
\delta U(\xi) &= -\xi U'(\xi) \nonumber \\
\delta P(\xi) &= (d+2-\omega)P(\xi)-\xi P'(\xi). \\
\end{align}
This can then be explicitly shown to solve the perturbation equations for $q=0$. This solution which is zero-order in q still does not allow us to determine the value of d at the small $\beta$ limit, as the divergence of the solution near the sonic point only appears at first order in q, and to that order the equations are not analytically tractable in general.


\begin{thebibliography}{99}
\bibitem{Sedov} L. I. Sedov, “Propagation of strong blast waves,” Prikl. Mat. Mekh. 10, 241 (1946)
\bibitem{Taylor} G. I. Taylor, “The formation of a blast wave by a very intense explosion,” Proc. R. Soc. London Ser. A 			201, 159 (1950).
\bibitem{VonNeumann} J. von Neumann, Blast Waves, Los Alamos Sci. Lab. Tech. Series (Los Alamos, NM, 1947), Vol. 7
\bibitem{WS} E. Waxman \& D. Shvarts, “Second-type self-similar solutions to the strong explosion problem“,
		Phys. Fluids A 5, 1035 (1993)
\bibitem{ryu} D. Ryu and E. T. Vishniac, “The growth of linear perturbations of adiabatic shock waves,” Astrophys. J. 				313, 820 (1987)
\bibitem{Kushnir} D. Kushnir, E. Waxman \& D. Shvarts, “The Stability of Decelerating Shocks Revisited“,  Astrophys J., 634, 407 (2005)
\bibitem{SWS} R. Sari, E. Waxman \& D. Shvarts, “Shock Wave Stability in Steep Density Gradients“, Astrophys. J. Sup., 				127, 475 (2000)
\bibitem{Landau} L. D. Landau and E. M. Lifschitz, Fluid Mechanics, 2nd ed. (Pergamon, New York, 1987)
\bibitem{Gruzinov} Gruzinov, A.\ 2003, arXiv:astro-ph/0303242 
\bibitem{BM} R. D. Blandford and C. F. McKee, “Fluid dynamics of relativistic blast waves“, Phys. Fluids 19, 1130 (1976)
\bibitem{best} P. Best and R. Sari, “Second-type self-similar solutions to the ultrarelativistic strong explosion 					problem“, Phys. Fluids 12, 3029 (2000)
\bibitem{sari} R. Sari, “First and second type self-similar solutions of implosions and explosions containing ultrarelativistic shocks“, Phys. Fluids 18, 027106 (2006)
\bibitem{pan} M. Pan \& R. Sari, “Self-Similar Solutions for Relativistic Shocks Emerging from Stars with Polytropic 				Envelopes“, Astrophys. J., 643, 416 (2006)



\end{thebibliography}
\end{document}